\documentstyle[aps,pra,graphicx,epsfig,multicol]{revtex}
\newcommand{\ea}{{\it et al. }}
\newcommand{\ie}{{\it i.e. }}

\newcommand{\0}{|0\rangle}
\newcommand{\1}{|1\rangle}

\begin{document}

\draft 
\title{Estimation of qubit states in a factorizing basis}
\author{Th. Hannemann, D. Rei\ss, Ch. Balzer,  W. Neuhauser, P. E. Toschek, and
Ch. Wunderlich}
\address{Institut f\"ur Laser-Physik, Universit\"at Hamburg, Jungiusstr. 9, 20355 Hamburg, Germany}
\date{4 October 2001}
\maketitle
\begin{abstract} 
The optimal estimation of a quantum mechanical 2-state system 
(qubit) - with $N$ identically prepared qubits available - 
is obtained by measuring all qubits simultaneously in an 
entangled basis. We report the experimental 
estimation of qubits using a {\em succession} of $N$ measurements on
{\em individual} qubits where  
the measurement basis is changed during the estimation procedure 
conditioned on the outcome of previous measurements 
(self-learning estimation). The performance of this adaptive 
algorithm is compared with other algorithms using measurements in 
a factorizing basis.
\end{abstract}
%Quantum computation

\begin{multicols}{2}

A question of fundamental and practical importance regarding the 
quantum mechanical description of the microscopic world is:
How can we obtain maximal information in order to characterize the 
state of a quantum system? Quantum states of various physical 
systems such as light fields, molecular wave packets, motional 
states of trapped ions and atomic beams have been determined 
experimentally with considerable precision \cite{review}. 
Acquiring complete knowledge about a quantum state would, of 
course, only be possible, if infinitely many copies of a quantum 
state were available and could be measured. More to the point, 
the initial question may be reformulated as the following task: 
Find a procedure consisting of a {\em finite} number of 
measurements yielding  a state vector that best represents the 
(classical) knowledge possibly gained from {\it any} type of 
measurement of the quantum system under scrutiny. 

Determining an arbitrary state of a quantum mechanical two-state 
system (qubit) is of particular importance in the context of 
quantum information processing. In Ref. \cite{Peres91} two 
identically prepared  2-state quantum systems were considered 
with no nonlocal correlations and it was searched for the optimal 
measurement strategy to gain maximal information   (difference of
Shannon entropy) 
about this quantum state. It was strongly suggested 
that optimal information gain is achieved when a suitable 
measurement on both particles together is performed. Later it was 
proven that, indeed the optimal measurement for determining a 
quantum state - if two spin 1/2 particles are available - needs 
to be carried out on both particles together, \ie the operator 
characterizing the measurement does not factorize into components 
that act in the Hilbert spaces of individual particles only 
\cite{Massar95}. Moreover, an optimal estimate of the spin 
direction (the qubit state) of an ensemble  of  $N$ identically 
prepared particles requires the application of such a 
nonfactorizable  measurement operator.  As a special case of 
optimal quantum state estimation of systems of arbitrary finite 
dimension the upper bound $(N+1)/(N+2)$ for the mean fidelity of 
an estimate of $N$ qubits was rederived in Ref. \cite{Derka98}. 
In particular, it was shown that {\it finite} positive operator 
valued measurements (POVMs) are sufficient for optimal state 
estimation.  This result implied that an experimental realization 
of such measurements is feasible, at least {\it in principle}. 
Subsequently, optimal POVMs were derived to determine the pure 
state  of a qubit with the {\it minimal} number of projectors  
when  up to $N=5$ copies of the unknown state are available  
\cite{Latorre98}. Still, the proposed optimal and minimal 
strategy requires the experimental implementation of rather 
intricate nonfactorizable operators for a simultaneous 
measurement on all $N$ qubits. First experimental steps towards 
entanglement-enhanced determination ($N=2$) of a quantum state 
have been undertaken \cite{Meyer01}. Estimating a quantum state 
can also be viewed as the decoding procedure at the receiver end 
of a quantum channel necessary to recover quantum information 
(e.g. encoded as a unit vector) \cite{Jones94,Bagan00}. 

It was recently shown that quantum state estimation with fidelity 
close to the optimum is possible when a self-learning algorithm 
is used and measurements on $N$ identically prepared qubits are 
performed separately, even successively \cite{Fischer00}. Here, 
we present, to our knowledge, the first experimental realization 
of a self-learning measurement on an individual quantum system in 
order to estimate its state. The base of measurement is varied in 
real time during a sequence of $N$ measurements conditioned on 
the results of previous measurements in this sequence. In 
addition, we compare the attainable experimental fidelity of this 
adaptive strategy for quantum state estimation with strategies 
where the measurement base is either a predetermined one, or is  
randomly chosen during a sequence of $N$ measurements. If a 
self-learning algorithm is employed to estimate a quantum state, 
then a suitable target function (here, the gain in the expected 
mean fidelity as described below) is maximized when proceeding 
from measurement $n-1$ to $n$. Under realistic experimental 
conditions possible errors have to be taken into account that may 
influence different measurement strategies differently. 

Here, the quantum mechanical two-state system under investigation 
is the S$_{1/2}$ ground-state hyperfine doublet with total 
angular momentum $F= 0,1$ of a single $^{171}$Yb$^+$ ion confined 
in a miniature Paul trap (diameter of 2 mm). The 
 \mbox{$\0 \equiv |F=0\rangle  \leftrightarrow  |F=1, m_F=0\rangle \equiv \1$} 
transition with Bohr frequency $\omega_0$ is driven by a 
quasiresonant microwave (mw) field with angular frequency near 
$\omega = 2\pi\,12.6$ GHz. The system is virtually free of 
decoherence, {\it i.e.} transversal and longitudinal relaxation 
rates are negligible  \cite{Huesmann99,Balzer01}. Photon-counting resonance 
fluorescence on the S$_{1/2}$(F=1) $\leftrightarrow$ 
P$_{1/2}$(F=0) transition driven by a frequency-doubled 
Ti:sapphire laser at 369~nm serves for state selective detection. 
Optical pumping into the $|F=1,m_F=\pm 1\rangle $ levels during a 
detection period is avoided when the \textit{E} vector of the 
linearly polarized light subtends 45$^o$ with the direction of 
the applied dc magnetic field. The light is detuned to the red 
side of the resonance line by some 20MHz in order to laser-cool 
the ion. Optical pumping the ion into the metastable $^2D_{3/2}$ 
level is prevented by illumination with light at 935~nm of a diode 
laser that retrieves the ion to the ground state via the 
$|D_{3/2}$, $F$=$1\rangle \rightarrow |[1/2]_{1/2}$, $F=0\rangle 
$ excitation. Cooling is achieved by simultaneously  irradiating 
the ion for 100~ms with light from both laser sources and with 
microwave radiation. This is done before each succession of 
measurements that consists of preparing and measuring a qubit 
state $N$ times. 

In the reference frame rotating with $\omega$, after applying the 
rotating wave approximation, the time evolution operator 
determining the evolution of the qubit exposed to linearly 
polarized mw radiation reads 
 $ U(t)= \exp\left[-\frac{i}{2}t\left( \delta
          \sigma_z + \Omega\sigma_x\right)\right]  $.
The Rabi frequency is denoted by $\Omega$ and $\sigma_{z,x}$ 
represent the usual Pauli matrices. 
Any pure state can be represented by a unit vector in 3D
configuration space (Bloch vector):
$|\theta,\phi\rangle=\cos\frac{\theta}{2}|0\rangle
 +\sin\frac{\theta}{2}e^{i\phi}|1\rangle$ and
is prepared by driving the qubit with mw pulses with 
appropriately chosen detuning $\delta \equiv \omega_0-\omega$, 
intensity, and duration, and by allowing for free precession for 
a prescribed time. Rabi frequency ($\Omega=3.47\times 2\pi$
kHz) and detuning  ($\delta=107\times 2\pi$ Hz) of the mw radiation
are determined by recording Rabi oscillations over 4-8 periods 
and by performing a Ramsey-type experiment with mw pulses 
separated in time. A measurement in a 
given direction is performed in two steps: First, a suitable 
unitary transformation of the qubit is performed effecting a 
rotation of the desired measurement axis onto the $z$-axis. 
Second, the qubit is irradiated for 2 ms with laser light 
resonant with the S$_{1/2}$(F=1) $\leftrightarrow$ P$_{1/2}$ 
transition and scattered photons are detected, if state $\1$ is 
occupied.

A self-learning measurement of the prepared qubit state consists 
of $N$ sequences each comprising i) the preparation of 
 $ |\theta_{\rm prep},\phi_{\rm prep}\rangle$, 
ii) performing a projective measurement in the basis 
 $\left(|\theta_{\rm m},\phi_{\rm m}\rangle_{\rm n}\,, 
|\bar{\theta}_{\rm m}\equiv\pi-\theta_{\rm m},
\bar{\phi}_{\rm m}\equiv\pi+\phi_{\rm m}\rangle_{\rm n}\right)$,  
and iii) using the result of this $(n-1)^{\rm th}$ measurement 
to determine the basis of the subsequent $n^{\rm th}$ measurement 
that maximizes the 
gain of the expected mean fidelity \cite{Fischer00}. This third 
step will be detailed in what follows. 

After $n-1$ sequences the density operator representing the 
state to be estimated is given by   
 $  \varrho_{n-1}=\int_0^\pi d\theta\sin\theta\int_0^{2\pi}d\phi
\;  w_{n-1}(\theta,\phi) \;|\theta,\phi\rangle\langle \theta,\phi|$.
The normalized probability density distribution 
$w_{n-1}(\theta,\phi)$ 
is updated after each measurement using Bayes rule \cite{Jones94}, 
\ie if in sequence $n$ the system is measured in direction  
  $(\theta_{\rm m},\phi_{\rm m})$ 
the distribution is modified by the probability for this outcome
  \begin{equation}
    w_n(\theta,\phi|\theta_{\rm m},\phi_{\rm m})=
    \frac{    w_{n-1}(\theta,\phi)\;
    |\langle \theta_{\rm m},\phi_{\rm m}|\theta,\phi\rangle|^2} {    
    p_{n}(\theta_{\rm m},\phi_{\rm m})}\; , 
  \label{eq:Bayes}
  \end{equation}
where the probability 
 $  p_n(\theta_{\rm m},\phi_{\rm m})=
    \langle\theta_{\rm m},\phi_{\rm m}|\varrho_{n-1}|\theta_{\rm
      m},\phi_{\rm m}\rangle 
 $ 
to find the system in direction 
$(\theta_{\rm m},\phi_{\rm m})$ 
in the $n-$th measurement ensures correct normalization.  

The best estimate of the pure qubit state, 
$|\theta_{\rm est},\phi_{\rm est}\rangle_{n-1}$ 
is obtained by maximizing the fidelity 
 $    F_{n-1}(\theta,\phi)= \langle\theta,\phi |\varrho_{n-1}
     |\theta,\phi\rangle  $, \ie 
 $ F_{n-1}(\theta_{\rm est},\phi_{\rm est})=F^{\rm opt}_{n-1}
 \equiv\max F_{n-1}(\theta,\phi)$. 
In order to find the optimal measurement direction for sequence 
$n$, the {\em expected} mean fidelity {\em after} measurement $n$ 
is maximized as a function of the measurement direction. Suppose 
in the $n$-th measurement the qubit is found in direction 
$(\theta_{\rm m},\phi_{\rm m})$. 
Then
   \begin{eqnarray}
     \lefteqn{F_n(\theta,\phi|\theta_{\rm m},\phi_{\rm m})=}
\nonumber\\
&&     \int_0^\pi \!d\theta'\sin\theta'\int_0^{2\pi}\!d\phi'
     \,  w_n(\theta',\phi'|\theta_{\rm m},\phi_{\rm m})
     \,|\langle\theta,\phi|\theta',\phi'\rangle|^2 \; ,
   \end{eqnarray}
where the expected distribution 
 $w_n(\theta',\phi'|\theta_{\rm m},\phi_{\rm m})$ 
is obtained from Bayes rule 
(eq.~\ref{eq:Bayes}). The optimal 
fidelity 
$F_n^{\rm opt}(\theta_{\rm m},\phi_{\rm m})$ 
is  
obtained by maximizing this function with respect to 
$(\theta,\phi)$. 
Measurement $n$ is performed along a specific  
axis and the qubit might as well be found in the direction 
$(\bar\theta_{\rm m},\bar\phi_{\rm m})$ 
Therefore, the expected 
mean fidelity after the $n-$th measurement is given by the  
optimized fidelities for each of the two possible outcomes, 
weighted with the estimated probability for that outcome:
   \begin{eqnarray}
     \bar F_n(\theta_{\rm m},\phi_{\rm m})  
     =&&p_n(\theta_{\rm m},\phi_{\rm m})
     F_n^{\rm opt}(\theta_{\rm m},\phi_{\rm m})\nonumber\\
     &+&\;p_n(\bar\theta_{\rm m},\bar\phi_{\rm m})
     F_n^{\rm opt}(\bar\theta_{\rm m},\bar\phi_{\rm m}) \; .
  \label{eq:F_opt}
   \end{eqnarray}
The optimal measurement direction 
 $(\theta_{\rm m}^{\rm opt},\phi_{\rm m}^{\rm opt})$
maximizes this function.

The direction of the first ($n=1$) measurement is of course 
arbitrary, since no {\it a priori} information on the state  
($w_0(\theta,\phi)=\frac1{4\pi}$) 
is available. The expected mean 
fidelity in this case is $\bar F_1=2/3$, independent of 
$(\theta_{\rm m},\phi_{\rm m})_1$. 
After the first measurement 
the symmetry of the probability distribution $w_1(\theta,\phi)$ is
reduced to rotational symmetry around the first 
measurement axis. 

\begin{figure}[htbp]
  \begin{center}
   \epsffile{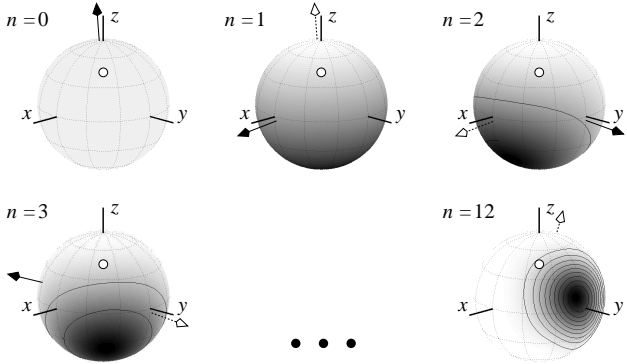}
    \caption{ 
      Probability distribution $w_n(\theta,\phi)$ on the Bloch
      sphere updated by Bayes rule, in a typical realization of 
      12 successive measurements. Darker areas correspond to higher
      probabilities (scaled individually for each Bloch sphere); 
      contour lines for $w_n=0,\;0.1,\;\ldots$ are given.
      The $n^{\rm th}$ and the optimized $(n+1)^{\rm th}$ measurement
      directions are indicated by the open and the solid arrows
      respectively. The white circle shows the prepared state
      $|\theta_{\rm prep},\phi_{\rm prep}\rangle=|\pi/4,\pi/4\rangle$.} 
    \label{fig1}
  \end{center}
\end{figure}

The expected mean fidelity now depends only on the relative 
angle $\alpha$ between the second and the first measurement 
direction and we find
  $\bar F_2=(1/2+\cos(\alpha/2-\pi/4)/\sqrt{18})$. 
Thus, the optimal second measurement with $\alpha =\pi/2$ yields 
$\bar F_2^{\rm opt}=(1/2+1/\sqrt{18})$. After the second 
measurement $w_2(\theta,\phi)$ 
is still symmetric with respect to 
a plane spanned by the first two measurements directions. Again, 
the optimal measurement direction axis is orthogonal to both 
previous directions and we obtain $\bar F_3^{\rm 
opt}=(1/2+1/\sqrt{12})$. The optimal directions of subsequent 
measurements ($n>3$) do depend on the outcome of previous 
measurements. For an estimation procedure comprised of $N$ 
sequences, we have calculated numerically $2^N$ possible 
successions of directions 
$\{(\theta_{\rm m},\phi_{\rm m})_n\}$ 
and programmed the computer interface that controls the 
experimental parameters to choose the optimum measurement 
direction online during an estimation procedure. Fig.~\ref{fig1} 
illustrates a succession of measurements that yield an estimate of the 
initial state 
$|\theta_{\rm prep},\phi_{\rm prep}\rangle =  |\pi/4,\pi/4\rangle$
employing the self-learning algorithm. The probability density 
$w_{n}(\theta,\phi)$ 
is shown on the surface of the 
Bloch sphere and the $n^{\rm th}$ and optimized
(eq.~\ref{eq:F_opt})  $(n+1)^{\rm th}$ 
measurement directions are indicated.

The discussion so far is based on the assumption, that  
measurements are performed with perfect efficiency. This is 
obviously not true in a real experiment. In this paragraph we 
will discuss the influence of experimental imperfections on the 
quality of state estimation.  Since the Rabi frequency $\Omega$ 
and detuning $\delta$ are determined precisely with an error 
below 1\%, the deviation of the prepared state and of the 
measurement axis from their anticipated directions is small and 
the resulting systematic error in the fidelity is negligible 
compared to the statistical error. If there were no background 
signal during a detection period, the observation of $m>0$ 
scattered photons in a single measurement would reveal the ion to 
be in state $\1$ with probability $1-p_1(0)$ close to unity. (The 
probability $p_{1}(m)$ to detect $m$ photons follows a Poissonian 
distribution with mean value $\bar m_1\approx5$.) However, due to 
scattering off the ion trap electrodes and windows some photons 
will be detected even if the ion had been prepared in state $\0$ 
(also with a Poissonian distribution $p_0(m)$ with   $\bar 
m_0\approx0.2$). In order to assign a given number of photon 
counts in an individual measurement to the corresponding state of 
the ion, the threshold $s$ is introduced: The probability 
$\eta_1$ to detect $m\geq s$ photons, when photons are scattered 
off the ion (state $\1$) is given by  $\eta_1=\sum_{m=s}^\infty 
p_1(m)$. Analogously, 
 $\eta_0=\sum_{m=0}^{s-1} p_0(m)$ 
for state $\0$. The functional relationship between $\eta_i$ and 
$s$ is determined by the observed photon number 
distributions $p_i(m)$. Since the detection efficiencies 
$\eta_i<1$, both a statistical and a systematic error are 
introduced into the measurements, as will be shown below. 

Using the {\em average} efficiency 
 $\bar\eta\equiv(\eta_0+\eta_1)/2$  
and the efficiency {\em difference} 
 $\Delta\eta\equiv (\eta_1-\eta_0)/2$,   
the probability to find an ``on" event ($m\geq s$) is given by   
 $P(\mbox{``on"})=(2\bar\eta-1)P_1+(1-\bar\eta)+\Delta\eta$,
and, analogously 
 $P(\mbox{``off\,"})=(2\bar\eta-1)P_0+(1-\bar\eta)-\Delta\eta$, where
$P_i=|\langle i|\Psi\rangle|^2$ and the $|\Psi\rangle$ is the 
ion's state before irradiation with UV light. This effect of the 
measurement can be thought of as the distorting action of a 
quantum channel  on the system's state followed by a perfect 
measurement:
 $    \varrho\to(2\bar\eta-1)
    \varrho+(1-\bar\eta)I+\Delta\eta\;\sigma_z 
 $
The channel acts as a depolarizing one characterized by the 
damping parameter $1-\bar\eta$. The error introduced hereby is 
independent of the choice of the measurement basis and hence  
statistical. Effectively the purity of the state (or equivalently 
the length of the Bloch vector $|\langle \vec \sigma\rangle|$) 
decreases. The term in the final density matrix containing 
$\Delta\eta$ systematically shifts the resulting state along the 
measurement direction. If an algorithm for state estimation is 
used that relies on measurements in fixed directions, for example 
in the $x$,- $y$- and $z$-direction, then the estimated state acquires 
a component parallel (or anti-parallel for $\Delta\eta < 0$) to 
the direction determined by the vector sum of the measurement 
directions. On the other hand, algorithms using measurement 
directions distributed over the whole Bloch sphere tend to cancel 
this error. This can be achieved with both the self-learning and 
the random algorithm. For experimental reasons we implemented 
only measurement directions on the upper hemisphere (\ie 
$\theta_{\rm m}\leq \pi/2$) and thus observe this systematic 
error for all algorithms if $\Delta\eta \neq 0$. Choosing the 
threshold $s_{\rm opt}$ such that $\Delta\eta = 0$ eliminates 
this systematic, basis dependent error. Whenever an efficiency 
difference cannot be avoided, any algorithm can be made   more 
robust against a systematic error in the state estimation by 
choosing measurement directions such that their vector sum is 
close to zero. Note that $\Delta\eta=0$ (no bias in the 
estimation procedure) does not, in general, occur at that 
threshold  that is required to make the most probable assignment 
to state $\0$ or $\1$ of a given number of detected photons. The 
threshold $s'_{\rm opt}$ required for the latter would be at the 
intersection of the two distributions, yielding the maximum of 
the detection efficiency $\bar{\eta}$. However, the photon count 
distributions in our experiment yield $s_{\rm opt}=s'_{\rm 
opt}=2$.

\begin{figure}[htbp]
  \begin{center}
   \epsffile{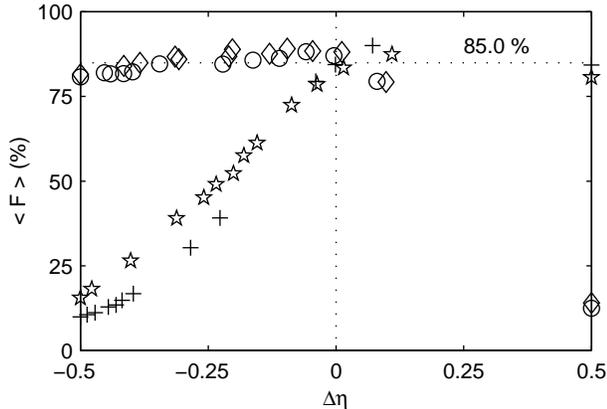}
    \caption{
      Mean fidelity attained with $N=12$ successive measurements,
      optimized by the self-learning algorithm, as a function of the 
      efficiency difference $\Delta\eta$ for different prepared states 
      (circles: $|3\pi/4,\pi/4\rangle$, diamonds: 
      $|3\pi/4,3\pi/4\rangle$, stars: $|\pi/4,3\pi/4\rangle$, plus: 
      $|\pi/4,\pi/4\rangle$). Each data point is averaged over 100-200 
      realizations.}
    \label{fig2}
  \end{center}
\end{figure}

We have studied the influence of the bias direction on the 
performance of all three algorithms. To this end  $\Delta\eta$ 
was varied by changing the threshold $s$ for the estimation of 
four different prepared states. Each state was estimated several 
hundred times after 12 consecutive measurements for a given value 
of $\Delta\eta$. Fig.~\ref{fig2} shows that the dependence of the 
fidelity on $\Delta\eta$ strongly varies for different states to 
be estimated. The curves in Fig. 2 intersect where the fidelity 
is independent of the prepared state. This intersection occurs at 
$\Delta\eta=0$ as is expected, if the functional dependence of 
$\Delta\eta$ on $s$ is correct (determined independently using 
the experimental photon count distributions.) 

In addition to the fidelity optimizing adaptive algorithm, an 
``orthogonal" and a ``random" one have been implemented for 
comparison. For the orthogonal algorithm equally many measurement 
are carried out in the $x$- $y$- and $z$- direction. The result 
of a succession of 4 measurements in each direction is evaluated 
by updating the probability distribution on the Bloch sphere 
using Bayes rule as is done for the adaptive algorithm. The 
random algorithm is realized by employing $2^N$ randomly 
generated directions instead of the $2^N$ optimized directions as 
described above for the self-learning measurement.

Table~\ref{table1} shows the fidelities for these three 
algorithms together with the respective values expected from 
theory.  The experimental fidelities for each algorithm are 
obtained at the intersection of four curves at $\Delta\eta=0$ 
corresponding to the estimation of  four different initial states 
as described above. The attainable fidelity is limited by 
experimental imperfections, \ie by the finite detection 
efficiency   $\bar\eta=97 \%$, and most notably by the impure 
preparation of state $\0$ at the beginning of each sequence of 
measurements ($\eta_{\rm prep}=89 \%$). The orthogonal algorithm 
in addition suffered from light-induced decoherence 
\cite{Balzer01}. These imperfections reduce the purity of the 
state, \ie the length of the Bloch vector 
$|\langle\vec\sigma\rangle|$. This is accounted for in the 
theoretical values (table~\ref{table1}), that are average values
obtained from numerically simulating each algorithm 10.000 times. It
should be emphasized that the fidelities given are average values  
valid for measurement sequences with $N=12$. If instead, the 
information of {\em all} sequences (typically $N\times 100=1200$) is used
for state esti\-mation, then the experimental  
fidelity is better than 99\%.

\begin{table}[htbp]
  \begin{center}
    \begin{tabular}{lccc}
      Algorithm
      &$\langle F \rangle_{\rm Exp.}$&$|\langle\vec\sigma\rangle|$ 
      &$ \langle F\rangle_{\rm Theo.}(|\langle\vec\sigma\rangle|)$\\\hline
      self-learning&$85.0\pm0.6$&$74.8\pm2.1$&$85.4\pm0.7$\\
      random&$81.9\pm0.6$&$73.4\pm2.1$&$81.9\pm0.7$\\
      orthogonal&$67.8\pm1.1$&$40.4\pm4.6$&$70.2\pm2.1$\\
      orthogonal&&$74.1\qquad\;\:$&$83.6\qquad\;\:$\\
    \end{tabular}
    \caption{Experimental mean fidelities (taken from the data plotted in
      Fig.~\ref{fig2} at: $\Delta\eta=0$), overall length of Bloch
      vector $|\langle\vec\sigma\rangle|$
      and mean fidelities expected from theory, when the length of the
      Bloch vector is taken into account. For comparison the expected
      value for the orthogonal algorithm is also given for a
      $|\langle\vec\sigma\rangle|$ as realized with the other algorithms.}  
    \label{table1}
  \end{center}
\end{table}

The method and results presented are not restricted to a 
particular realization of qubits. 
The identification of imperfections in our experiment show that the
fidelities obtained are currently limited
mainly by the impure preparation. Here, a
significant improvement seems feasible:
$\eta_{\rm prep}\gtrsim 99\%$ 
would lead to mean fidelities (with $N=12$ qubits)
better than $90 \%$, close to the
upper bound of $93 \%$, attainable with an entangled measurement.

This work was supported by the Deutsche Forschungsgemeinschaft 
and the Bundesministerium f\"ur Forschung und Technologie.

\end{multicols}

\end{document}